\documentclass[10pt,aps,prx,twocolumn,showpacs,notitlepage,floatfix,superscriptaddress,longbibliography]{revtex4-2}
\usepackage{hyperref}
\hypersetup{
colorlinks=false,
colorlinks=true,
citecolor=blue,
linkcolor=blue,
urlcolor=blue}
\usepackage{graphicx}
\usepackage{amsmath}
\usepackage{physics}
\usepackage{amsfonts}
\usepackage{amssymb}
\usepackage{mathtools}
\usepackage{bm}
\usepackage{siunitx}
\usepackage{xfrac}
\usepackage[normalem]{ulem}
\usepackage{orcidlink}  
\usepackage{braket}
\usepackage{cases}

\newcommand{\mdas}[1]{\textcolor{black}{#1}}
\newcommand{\pcmd}[1]{\textcolor{black}{#1}}

\DeclareUnicodeCharacter{0229}{\c{e}}

\begin{document}

\title{Exploring the entropic asymmetry on logical stochastic resonance with energetically equivalent intrinsic outputs}
\author{Vipul Rai\orcidlink{0009-0000-0813-4482}}
\email{Contact author: d23176@students.iitmandi.ac.in}
\affiliation{School of Physical Sciences, Indian Institute of Technology Mandi, Mandi-175005 (H.P.), India}
\author{Moupriya Das\orcidlink{0000-0003-1851-2162}}
\email{Contact author: moupriya@iitmandi.ac.in}
\affiliation{School of Chemical Sciences, Indian Institute of Technology Mandi, Mandi-175005 (H.P.), India}

\begin{abstract}

Small-scale systems \mdas{are inherently subject to environmental noise that} can be harnessed 
constructively to realize reliable logic operations --- a phenomenon known 
as logical stochastic resonance (LSR), where a bistable system produces  
correct logical outputs within an optimal window of noise intensity. 
The Brownian dynamics governed by appropriate inputs inside a double-well potential, modeling the bistable system, mimic the logic operations. The two wells of this potential represent two distinct logical output states $0$ and $1$.
Asymmetry in this potential is known to be essential for improving 
logical reliability. However, prior studies have focused on energetic 
asymmetry, characterized by unequal depths of the two wells of the potential.
This left the role of the width asymmetry in the potential, unexplored. This latter class of asymmetry emerges due to the dissimilar widths of the two wells of the potential. It can be classified as the entropic asymmetry between the two logical output states. Here, we systematically 
investigate the effect of width or the entropic asymmetry in the system on the logical response for OR 
and AND gate operations. Unlike energetic asymmetry, width asymmetry 
preserves the energetic equivalence of the two intrinsic logical output states, making it a geometric effect. We find that increasing width asymmetry 
consistently improves the optimal  $P_{\mathrm{logic}}$, the quantifier measuring the successful logical outcome.
Moreover, when it is combined with an 
energetic bias, it produces reliable logic gate operation at a significantly 
reduced energetic cost compared to the symmetric case. 
The requirement of this energy bias also diminishes gradually with the increasing degree of width asymmetry in the potential.\\ 







\end{abstract}
\maketitle

\section{Introduction}
The operations underlying modern digital computing are built upon binary 
logic, where information is encoded in one of two discrete states: $0$ or $1$. 
The physical implementation of these logical operations is realized 
through logic gates, which form the elementary building blocks of all 
digital circuits. As technological progress continues to drive 
miniaturization of devices toward the nanoscale, thermal fluctuations 
become increasingly significant and can no longer be neglected~\cite{KISH2002144,sano2000increasing,bulsara2005no}. At such 
scales, the deterministic picture of classical logic breaks down, and the 
interplay between logical signals, noise, and the nonlinear dynamics of 
the system must be understood at the microscopic level. 

A particularly intriguing consequence of operating in this noisy regime 
is the phenomenon of stochastic resonance (SR)~\cite{RevModPhys.70.223,Wiesenfeld_chaos,BULSARA2010424,Peter_prl_2003,peter_pra_1991,Adi_pre_2000,Casado-Pascual_pre,gammaitoni1993stochastic}, wherein an optimal 
level of noise can enhance, rather than degrade, the response of a 
nonlinear system. This counterintuitive effect --- that noise, usually 
regarded as a source of disturbance, can in fact improve the performance 
of both natural and artificial systems~\cite{Kish_APL_2006} --- has attracted considerable 
attention across a wide range of disciplines~\cite{Gammaitoni,Aharonov,RevModPhys.70.223,PhysRevLett.95.080503,Nori_prl,Nori_pre, Gammaitoni_prl,Astumian,Worschech,Hafiz,Li_prl}. Building on this 
insight, the concept of logical stochastic resonance (LSR) was introduced 
by Murali et al.~\cite{Murali_prl_2009,PhysRevA.39.4854,Murali_apl_2009,Sinha_2009,KOHAR2012957,Hou2020,Sinha_pre_2011}, who demonstrated that a \pcmd{two-state} system driven 
by two aperiodic input signals can produce correct logical outputs within 
an optimal window of noise intensity. 
\mdas{Here, a double-well potential representing the essential bistability \pcmd{of the two-state system} serves as the 
physical model of the system performing logic operations. The two wells of the 
double-well potential correspond to two distinct binary memory states $0$ and $1$.} 
Furthermore, structural asymmetry \mdas{arising from different depths of the two wells} 
of the \mdas{effective} bistable potential has been shown to play an important 
role in enhancing logical reliability, enabling more robust logic gate operation 
in the presence of noise~\cite{Murali_prl_2009,YU_PhysRevE_2023,Das_PRE,Das_pre_2012,Das_pre_2013,Murli_pra_2025,Zamora-Munt:10,CHENG2020109514,Dong_pre_2026,Parmananda_pre_2021,Gui_chaos,Sinha_pre_2011}.
Motivated by these developments, researchers have investigated the 
implementation of fundamental logic gates --- such as OR, AND, NAND, and 
NOR --- within the LSR framework~\cite{Murali_prl_2009, YU_PhysRevE_2023,Chen_pre_2026}. 
More recently, this line of work has been extended to realize XOR and XNOR 
gates~\cite{Cai_2025}, demonstrating the versatility of noise-assisted 
logic in bistable systems.

In the present work, we focus on an asymmetric bistable system in which 
the asymmetry arises solely from a difference in well widths \mdas{of the model double-well potential}, referred to 
here as a width or entropic asymmetry. Such a width asymmetry has been studied 
in various physical contexts, including the mean first-passage 
time~\cite{ciampini2021experimental, innerbichler_2020,Chupeau_2020,QIAO20212194}, memory 
erasure~\cite{Rai_vipul, Bechhoefer_prl}, and dynamical 
hysteresis~\cite{ghosh2025dynamic}, where it has been shown to offer 
distinct advantages over the symmetric case. 
\mdas{In physical devices, the width asymmetries in the potential are particularly observed in MEMS (Micro-Electro-Mechanical Systems), NEMS (Nanoelectromechanical systems), magnetic bistable systems with shape anisotropy, ferroelectric devices with defects, etc.}
\textcolor{black}{From an experimental 
standpoint, potential barrier shaping has recently been 
realized in MEMS (Micro-Electro-Mechanical Systems) devices. Here, tuning the electrostatic potential 
landscape has been shown to improve noise-assisted signal detection 
under varying noise conditions~\cite{MEMS_SR_2026}, demonstrating 
that geometric modification of bistable potentials is not merely a 
theoretical tool but an experimentally viable strategy.} 

Inspired by these findings, 
we investigate whether width asymmetry can similarly enhance the 
reliability of logical operations in the LSR framework. 
\mdas{While we understand that the energetic or depth-asymmetry in the underlying potential 
can improve the reliability of the logic response, we aim to explore this aspect for the 
bistable systems with entropic or width-asymmetric origin. 
The idea is to provide a comparative interpretation of the outcomes in these two scenarios.}

\mdas{We indicate one important point here regarding the energy values of the binary memory states. 
In the case of the depth-asymmetric system performing logic operations, 
the energy values of the memory states are different even when the systems 
are not subject to any inputs. \pcmd{This can be understood from the description of the analog circuit that was introduced to explain LSR~\cite{Murali_apl_2009,BULSARA2010424}. Here, considering two separate thresholds, or a DC bias, or both simultaneously, makes the effective double-well potential asymmetric; the two wells of the potential differ in depth.} Therefore, the exact energy \pcmd{of the intrinsic binary states $0$ and $1$, becomes distinct as each of them is represented by the state in a} particular well of the double-well potential. 
Whereas for the systems with width-asymmetry, the binary memory states are maintained 
at the same energy level in the static condition. Therefore, we can say that 
for these kinds of systems, we examine the logical response for 
energetically equivalent memory states to begin with. 
The findings establish the distinctions between the characteristics of the 
logic functions for these two specific classes of asymmetric systems.}

The paper is organized as follows. In Section II, the model 
and dynamics are described. \mdas{The logic gates implementation is defined in Section III.} 
The results have been presented and discussed in Section IV. Conclusions are presented in Section V.
 
\section{Model and Dynamics}

We model the logical operation \mdas{conventionally}~\cite{Murali_prl_2009,Murali_apl_2009,Zhang_pre,Das_PRE,YU_PhysRevE_2023,Guerra_nano_letters} through the dynamics of a 
Brownian particle confined to a one-dimensional bistable potential, subject to two external 
input signals. The dynamics of the particle are governed by the overdamped 
Langevin equation:
\begin{equation}
    \gamma \frac{dx}{dt} = -\frac{\partial V(x)}{\partial x} + I_1(t) + I_2(t) 
    + \sqrt{D}\, \xi(t),
    \label{eq:langevin}
\end{equation}
where $x(t)$ denotes the position of the Brownian particle, whose value 
determines the logical \mdas{output} state of the system. $\gamma$ is the friction 
(damping) coefficient. $\xi(t)$ is Gaussian white noise satisfying the 
statistical properties
\[
\langle \xi(t) \rangle = 0, \quad \langle \xi(t)\,\xi(t') \rangle = 
2\delta(t - t').
\]
\mdas{The noise term accounts for the environmental fluctuations which are common 
in logic devices of small scale.} $D$ is the diffusion coefficient, 
which serves as a measure of \mdas{the strength or magnitude of the} thermal fluctuations.   
It is related to the temperature $T$ and the damping 
coefficient $\gamma$ through \mdas{the relation} $D = \gamma k_{B}T$, where $k_{B}$ is the 
Boltzmann constant. $I_1(t)$ and $I_2(t)$ represent two external input 
signals that encode the logical inputs. Each signal takes one of two 
discrete values, corresponding to logical $0$ or logical $1$, and together 
they determine the net bias applied to the particle. 

\mdas{We consider two external signals, as at least two input signals are required to construct  
the OR, AND or NOR, NAND gates, the basic types of logic gates, which is the central goal of 
the present study. It is evident from the Langevin dynamics model represented through the 
Eq.~\eqref{eq:langevin}, the dynamical variable gets directly influenced by the action 
of the inputs. Therefore, it is justified to consider the state variable $x$ to determine the 
output which results from the operations of the inputs. Its value is controlled by the underlying 
potential $V(x)$ as well, which is generally considered to be of a double-well form. 
This is because at least two distinct states are required to represent the 
output values $0$ and $1$ emerging from the binary logic operation. 
The double-well potential serves as the minimal model to correspond to this two-state system 
which can mimic the intended logic operations. The state of the Brownian particle 
in two different wells is assigned to a specific logical output value $0$ or $1$. 
For example, when the particle is in the left well, it can be considered to denote 
the output state $0$, and in that case, its presence in the right well will signify 
the logical output $1$. The reverse convention can also be taken into account.   
A particular fashion of this assignment is followed depending on the type of logic gates that are 
planned to be designed. }

\mdas{It has been reported that it is important to incorporate asymmetry in 
this double-well potential model to obtain reliable logical responses.~\cite{Murali_prl_2009,Murali_apl_2009} 
In these previous studies, the asymmetry in the potential appears in terms of the different depths 
of the two potential wells, making the two logical output states energetically different.} 
\mdas{Here,} we use an asymmetric double-well potential $V(x)$ 
following the form used in Refs.~\cite{innerbichler_2020}, where the 
asymmetry arises from differing well widths \mdas{and shape of the barrier separating the 
two wells. This form of asymmetry in the potential is recognizable from the asymmetric structures 
considered earlier, as here the logical output states are energetically equivalent 
to start with.}. 

The potential is defined as
\begin{align}\label{eq:potential}
    V(x) &= h \left[ 1 + \left( \frac{2x}{S(x)} \right)^4 
    - 2\left( \frac{2x}{S(x)} \right)^2 \right], \\
    \textrm{with} \nonumber \\
    S(x) &= c\,\Theta(-x) + (2 - c)\,\Theta(x),\nonumber
\end{align}
where $h$ denotes the barrier height, $S(x)$ is a position-dependent 
scaling function that controls the relative widths of the left and right 
wells, and $\Theta(x)$ is the Heaviside step function. The positive 
parameter $c$ introduces structural asymmetry into the potential by 
setting the width ratio between the two wells. Unlike the erasure protocol 
considered in Refs.~\cite{Rai_vipul}, 
the potential here remains 
static throughout the operation; no barrier modulation is applied~\cite{innerbichler_2020}. The 
logical output is instead determined entirely by the net bias imposed by 
the input signals $I_1(t)$ and $I_2(t)$. 

We now make Eq.~\eqref{eq:langevin} into dimensionless form to work with 
a simplified description of the dynamics. The position is scaled as 
$\tilde{x} = x/l$, where $l$ is the distance between the two minima of 
the potential, with numerical value equal to $1$. Time is scaled by the 
factor $t_l = \gamma l^2 / k_B T_R$, where $T_R$ is a reference 
temperature, so that the dimensionless time is $\tilde{t} = t/t_l$. 
The factor $t_l$ effectively represents twice the time required for the 
Brownian particle to diffuse across the distance $l$. 
The input signals are made dimensionless by scaling with the thermal energy 
per unit length as $\tilde{I}_{1,2} = I_{1,2}\, l/k_B T_R$.
The dimensionless Langevin equation then takes the form
\begin{equation}
    \frac{d\tilde{x}}{d\tilde{t}} = 16h \left\{ 
    \frac{\tilde{x}}{S(\tilde{x})^2} - 4\frac{\tilde{x}^3}{S(\tilde{x})^4} 
    \right\} + \tilde{I}_1 + \tilde{I}_2 
    + \sqrt{\tilde{D}}\,\tilde{\xi}(\tilde{t}),
    \label{eq:langevin_dimless}
\end{equation}
where $\tilde{\xi}(\tilde{t})$ is the rescaled noise term, $\tilde{D} = 
T/T_R$ is the dimensionless diffusion coefficient, and $h$ is the 
dimensionless barrier height. In the following, we drop the tildes for 
brevity, with all quantities understood to be dimensionless. 

\begin{figure*}[ht]
    \includegraphics[width=\linewidth,height=4.5cm]{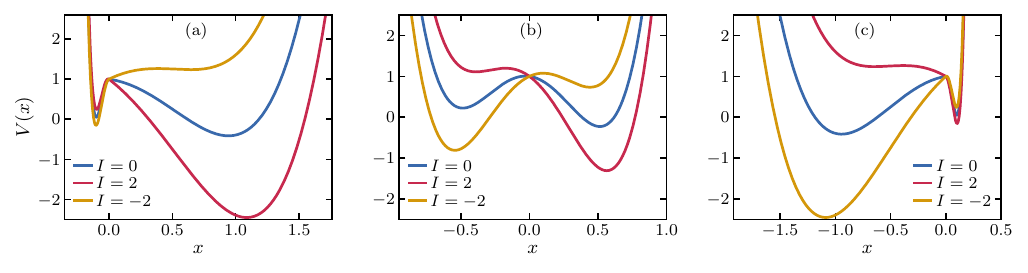}
    \caption[]{Asymmetric bistable potential \textbf{(a)} configuration for OR gate for \(c=0.2\), \textbf{(b)} OR gate configuration for \(c=1\) and \textbf{(c)} AND gate configuration for \(c=1.8\)}
  \label{fig:potential}
\end{figure*}

\begin{table*}[ht]
\centering
\caption{Truth tables for the OR and AND logic gates showing 
the logical output for each input combination $(I_1, I_2)$.}

\begin{tabular*}{\textwidth}{@{\extracolsep{\fill}}ccccc@{}}
\hline\hline
$I_{\mathrm{tot} \,= \, I_1 +I_2}$ 
&  $I_1$ and $I_2$ 
& Logical input sets 
& OR 
& AND \\ 
\hline

$-2$ 
& $(-1,-1)$ 
& $(0,0)$ 
& $0 \; (x < 0)$ 
& $0 \; (x < 0)$ \\

$0$ 
& $(-1,1)/(1,-1)$ 
& $(0,1)/(1,0)$ 
& $1 \; (x > 0)$ 
& $0 \; (x < 0)$ \\

$2$ 
& $(1,1)$ 
& $(1,1)$ 
& $1 \; (x > 0)$ 
& $1 \; (x > 0)$ \\

\hline\hline
\end{tabular*}

\label{tab:OR_AND}
\end{table*}

\begin{figure*}[ht]
  \centering
    \centering
    \includegraphics[width=0.92\linewidth,height=4cm]{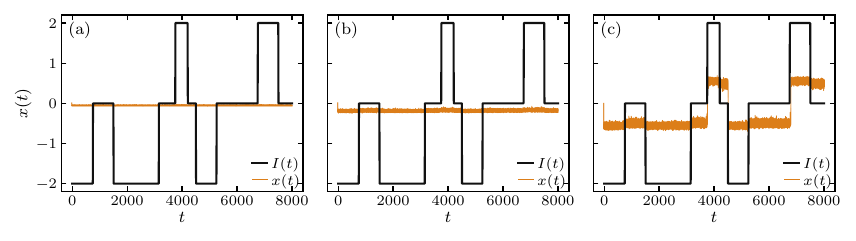}
    \label{fig:traj_c1}
 
  \vspace{0.1cm}
    \centering
    \includegraphics[width=0.92\linewidth,height=4cm]{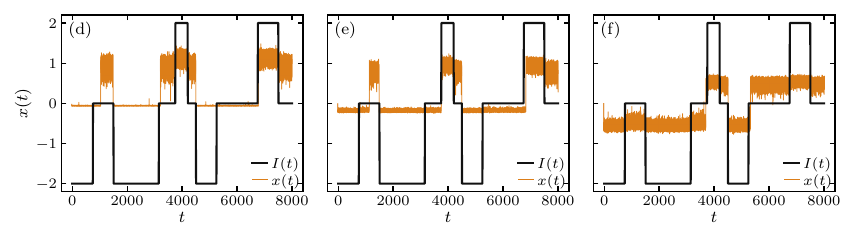}
    \label{fig:traj_c02}
    \vspace{0.1cm}
    \centering
    \includegraphics[width=0.92\linewidth,height=4cm]{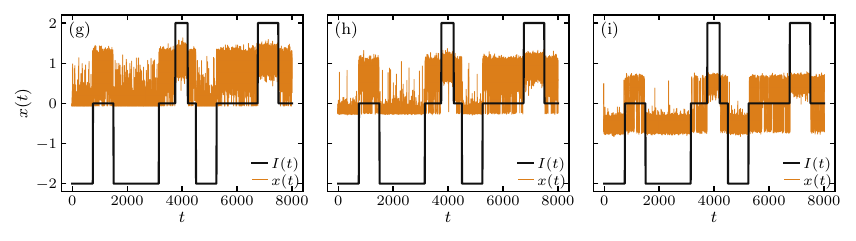}
    \label{fig:trajectories_5050}
    \caption{Input--output correspondence of the OR gate for different values of $D$ and $c$. The rows correspond to $D=0.05$, $D=0.125$, and $D=0.25$, respectively, while the columns correspond to $c=0.08$, $c=0.34$, and $c=1$, respectively.}
  \label{fig:trajectories}
\end{figure*}


\begin{figure*}[ht]
  \centering
    \centering
    \includegraphics[width=0.92\linewidth,height=4cm]{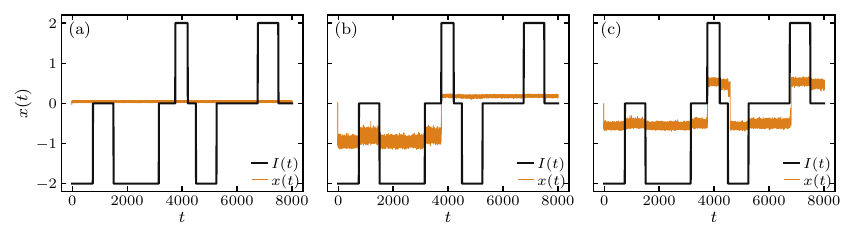}
 
  \vspace{0.1cm}
    \centering
    \includegraphics[width=0.92\linewidth,height=4cm]{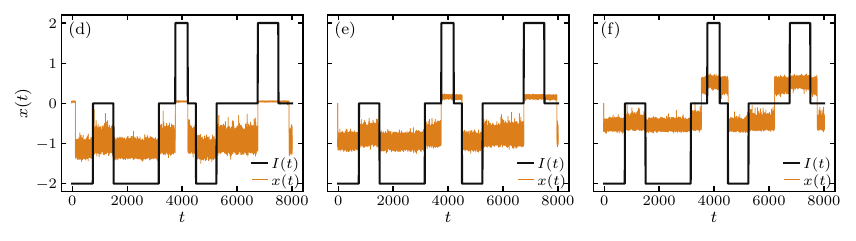}
    \vspace{0.1cm}
    \centering
    \includegraphics[width=0.92\linewidth,height=4cm]{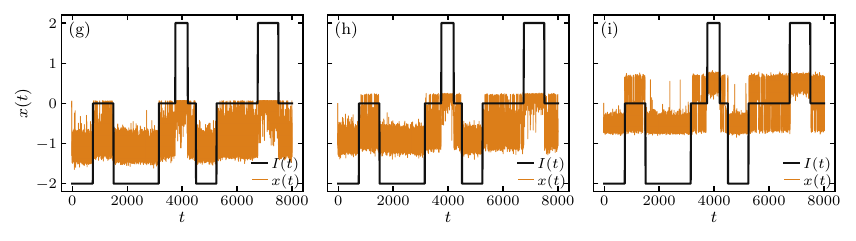}
    \caption{Input--output correspondence of the AND gate for different values of $D$ and $c$. The rows correspond to $D=0.05$, $D=0.125$, and $D=0.25$, respectively, while the columns correspond to $c=1.92$, $c=1.66$, and $c=1.00$, respectively.}
  \label{fig:trajectories_AND}
\end{figure*}

\section{Logic Gate Implementation}

We discuss the realization of OR and AND logic operations in an asymmetric 
bistable potential driven by noise. In the binary framework underlying 
digital computation, information is encoded in two discrete 
states: ON and OFF, \mdas{or TRUE and FALSE. The ON (or TRUE) 
and OFF (or FALSE) states are} assigned the logical values $1$ and $0$, respectively. \mdas{Both the inputs and outputs are interpreted with this binary representation.}

The two logical inputs $I_1$ and $I_2$ each take the value $+1$ 
(logical $1$) or $-1$ (logical $0$). Their superposition forms a 
three-level composite signal $I = I_1 + I_2$, which takes the values 
$\{-2,\, 0,\, +2\}$ corresponding to the four input combinations 
$(0,0)$, $(0,1)$\,/\,$(1,0)$, and $(1,1)$, respectively. The net signal 
$I$ thus acts as the driving force that biases the particle toward one 
of the two wells \mdas{of the bistable potential Fig.~\ref{fig:potential}}. 
\mdas{As the occupancy of a particular well occurs as 
a result of the action of the inputs, the state of the 
particle in the given well is considered as the output.}
While the bistable potential possesses two wells, the left 
well is associated with the logical \mdas{output} state $0$ 
and the right well with the logical \mdas{output} state $1$. 
The logical input--output correspondence for the OR 
and AND gates is summarized in Table~\ref{tab:OR_AND}. 

The width asymmetry of the potential is controlled through the parameter 
$c$. For $c < 1$, the left well is narrower than the right, and the 
system realizes an OR gate, whereas for $1 < c \leq 2$, the right well 
is narrower, and the system realizes an AND gate \mdas{under optimal control over the noise-strength}. The symmetric case 
$c = 1$ serves as the reference configuration. The primary objective 
of this work is to systematically investigate how the degree of width 
asymmetry, modulated by $c$, influences the reliability of these logical 
operations in the presence of noise. 

\mdas{Here, we point out that it is possible to consider the 
alternative definitions of the logical output states. i.e., 
the left well can be mapped to the output state $1$ 
and the right well to $0$. In that case, the systems will 
perform as an NOR and NAND gate within the given framework.}

\section{Numerical results and discussions}
\subsection{Input-output correspondence}
We consider the overdamped Langevin equation~(\ref{eq:langevin_dimless}) and solve it numerically using the improved Euler algorithm~\cite{press1992numerical}. The Gaussian white noise term $\xi(t)$ is generated using the Box--Muller algorithm. The time step used in the simulation is $\Delta t = 10^{-4}$. The input signals $I_1(t)$ and $I_2(t)$ are two aperiodic signals incorporated into the Langevin dynamics. \mdas{They appear in the form of the time-dependent square wave functions, which can have only two discrete levels. We consider these discrete magnitudes to have numerical values $1$ and $-1$, which correspond to the logical input values $1$ and $0$, respectively. Therefore,} the combination $I_1=-1$ and $I_2=-1$ indicates the logical input $(0,0)$, while $I_1=-1$ and $I_2=1$, or vice versa, signify the logical inputs $(0,1)/(1,0)$. Similarly, $I_1=1$ and $I_2=1$ corresponds to the logical input $(1,1)$, as listed in Table~\ref{tab:OR_AND}. 

We examine the input--output correspondence for different values of the asymmetry parameter $c$ \mdas{for the OR logic gates first}. To start with, we consider the low-noise case with $D=0.05$ for different asymmetry values, namely $c=0.20$, $0.56$, and $1$, as shown in Fig.~\ref{fig:trajectories}{\textcolor{blue}{(a)--(c)}}. In this low-noise regime, the system does not exhibit a proper logical response. When the noise strength is increased such that the barrier-height-to-noise ratio, $\Delta U/ k_B T$, 
reaches a suitable level, for example at $D=0.125$, the system begins to show a clear logical response, as illustrated in Fig.~\ref{fig:trajectories}{\textcolor{blue}{(d)--(f)}}. However, upon further increasing the noise strength to $D=0.25$, shown in Fig.~\ref{fig:trajectories}{\textcolor{blue}{(g)--(i)}}, the logical response again disappears due to excessive fluctuations induced by strong noise. Thus, at high noise strengths, random fluctuations dominate the dynamics, leading to unreliable logical responses. Therefore, the system exhibits optimal logical behavior only within an intermediate range of noise intensity. 

\mdas{The same analyses have been done in the context of realizing AND logical behavior exhibited by the systems. The results for the $c$ values $ c =1.92$, $1.56$ and $1.20$ at three different noise levels have been illustrated in Fig.~\ref{fig:trajectories_AND}{\textcolor{blue}{(a)--(i)}}. Here, the outcomes of the logical response also show a similar dependence on the noise strength, as discussed for the OR logic systems.}

\subsection{Quantifying the Logical Response}

To quantify the reliability of the logical response in a noisy bistable 
system, we compute the probability of logical success $P_{\mathrm{logic}}$, 
estimated from repeated stochastic realizations of the dynamics under all 
possible combinations of logical inputs.

For each logical input combination, the system is first evolved for a 
finite thermalization period $T_{\mathrm{settle}}$. This allows the particle, \mdas{the position of which represents the state of the system,}
to approach a steady state before any measurement is performed. This step 
excludes transient effects from the logical response, since the particle's 
initial position may not be consistent with the applied input.

Following the thermalization period, the system is evolved for an 
additional measurement interval $T_{\mathrm{measure}}$, during which the 
well occupation is monitored. We define the time-averaged occupation 
fraction of the right well as
\begin{equation}
    f_+ = \frac{t_+}{T_{\mathrm{measure}}},
    \label{eq:time_frac}
\end{equation}
where $t_+$ is the total time spent by the particle in the right well 
$(x > 0)$. A value of $f_+ \simeq 1$ indicates predominant occupation 
of the right well, while $f_+ \simeq 0$ indicates predominant occupation 
of the left well.

Rather than assigning the logical output from a single instantaneous 
position of the particle, we employ a fraction-of-time occupation 
criterion. A threshold $\theta$ is introduced, and the logical output 
is assigned as
\begin{equation}
    \text{Output} =
    \begin{cases}
        1 & \text{if} \quad f_+ > \theta, \\
        0 & \text{if} \quad f_+ < 1 - \theta.
    \end{cases}
    \label{eq:out}
\end{equation}
If $1 - \theta \leq f_+ \leq \theta$, the trajectory is regarded as 
ambiguous (constantly switching) and the trial is counted as unsuccessful. We set $\theta = 0.98$ 
throughout this work. This criterion ensures that a logical output is 
accepted only when the particle remains in the corresponding well for 
nearly the entire observation interval. This verifies not only correct 
switching behaviour but also the temporal stability of the logical state 
against noise-induced fluctuations.

The four logical input combinations are applied sequentially in a randomly 
chosen order within each trial. A trial is considered successful only if 
all four input combinations yield the correct logical output according to 
the chosen truth table. The probability of logical success is defined as~\cite{Murali_prl_2009,Zhang_pre,Cai_2025}
\begin{equation}
    P_{\mathrm{logic}} = \frac{N_{\mathrm{success}}}{N_{\mathrm{trial}}},
    \label{eq:P_logic}
\end{equation}
where $N_{\mathrm{success}}$ is the number of successful trials out of 
$N_{\mathrm{trial}}$ total realizations.

\subsubsection*{Entropic Bias}

\begin{figure}[ht]
\centering
\begin{minipage}{0.49\linewidth}
    \centering
    \includegraphics[width=\linewidth,height=3.5cm]%
    {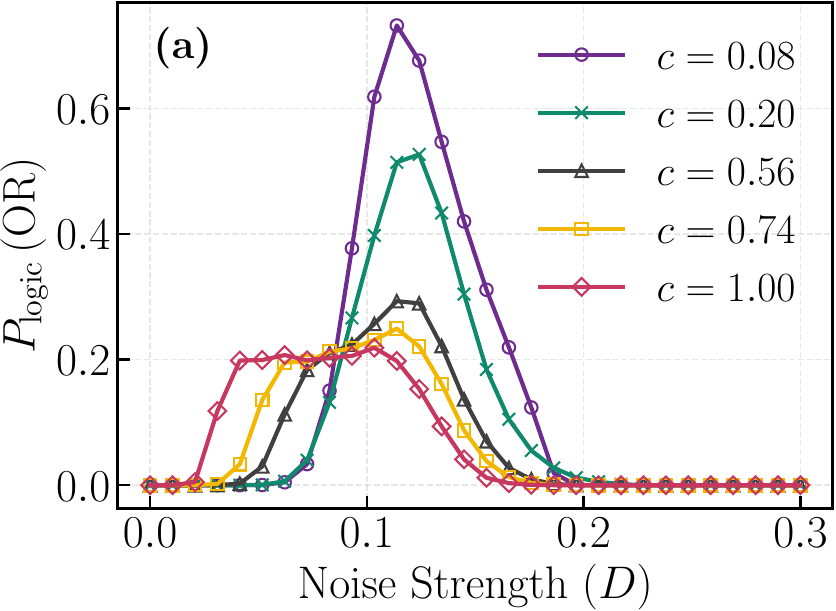}
\end{minipage}
\hfill
\begin{minipage}{0.49\linewidth}
    \centering
    \includegraphics[width=\linewidth,height=3.5cm]%
    {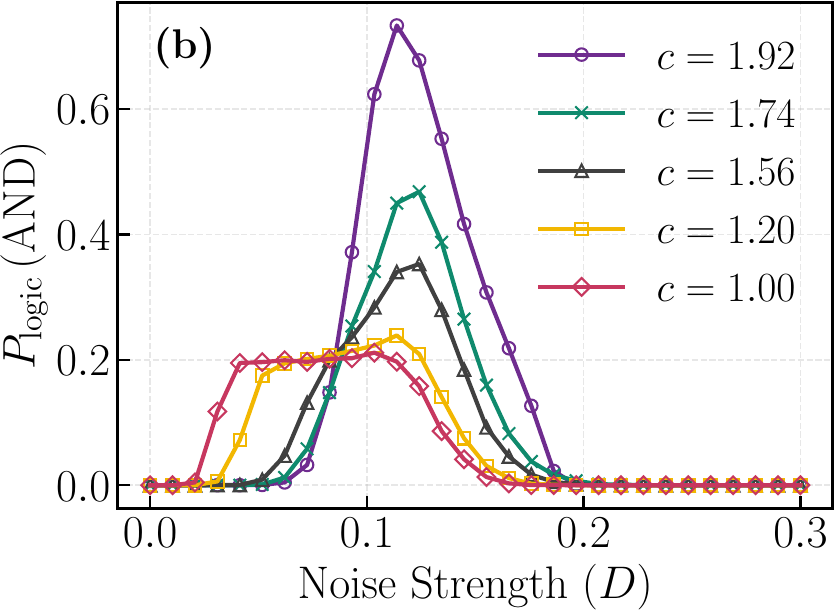}
\end{minipage}
\caption{Probability of logical success $P_{\mathrm{logic}}$ as a 
function of noise intensity $D$ for the (a) OR gate and (b) AND gate, 
evaluated over $N_{\mathrm{trial}} = 2 \times10^4$ trials. Results are shown 
for a purely entropic potential ($b = 0$) with varying width asymmetry 
parameter $c$. The thermalization and measurement intervals are 
$T_{\mathrm{settle}} = T_{\mathrm{measure}} = 300$, with occupation 
threshold $\theta = 0.98$.}
\label{fig:logical_response}
\end{figure}

We first consider a \mdas{bistable} potential with width asymmetry \mdas{between the two wells}. \mdas{The potential has the mathematical form represented by} Eq.~\ref{eq:potential}. 
\mdas{This potential can be considered to have pure entropic bias between the two wells with} no energy bias. The \mdas{quantification of the} logical response for the OR and AND gates is shown in Fig.~\ref{fig:logical_response}~{\textcolor{blue}{(a)--(b)}}, 
\mdas{respectively, as a function of noise strength $D$}. \mdas{The quantifiers $P_{\mathrm{logic}}(OR)$ and $P_{\mathrm{logic}}(AND)$ are} evaluated over $N_{\mathrm{trial}} = 2 \times10^4$ trials. 

\mdas{The study has been performed for a range of asymmetric systems with a varied degree of 
asymmetry in the underlying potential. For all cases, $P_{\mathrm{logic}}$ exhibits a maximum 
value at an intermediate magnitude of the noise strength $D$ or within a given span of it. 
This dependence on $D$ is justified by the examination of the input-output correspondence at 
different noise levels discussed in the previous subsection. This variation also connects to the 
similar observations described previously as logical stochastic resonance~\cite{Murali_prl_2009,Das_pre_2013,YU_PhysRevE_2023,Cai_2025,Dong_pre_2026}, where 
maximum reliable logical response is observed at an intermediate level of noise strength. 
However, our findings are distinct from the previous studies where asymmetry was considered to arise due to the different depths of the two wells of the double-well potential governing the dynamics.} 

\mdas{The main objective of the present work is to analyze the effect of entropic asymmetry on the
logical performance of the bistable system, modeled with the described double-well potential. 
Before discussing the systematic understanding of the degree of asymmetry in the systems 
on their logical performance, we first elucidate some important characteristic behavior of 
the symmetric system as a reference point.}
\textcolor{black}{A notable feature of Fig.~\ref{fig:logical_response} is the emergence of 
a flat region, or plateau, in $P_{\mathrm{logic}}$ at intermediate noise 
intensities as the system approaches the symmetric configuration ($c = 1$). 
This behaviour can be understood by examining the OR gate 
(Fig.~\ref{fig:logical_response}~{\textcolor{blue}{(a)}}) as a representative 
case. For $c = 1$, the potential is perfectly symmetric, with both wells 
having equal width and depth. The problematic input combinations for the 
OR gate are $(0,1)$ and $(1,0)$, which correspond to a vanishing net 
signal $I_{\mathrm{tot}} = I_1 + I_2 = 0$, yet the correct logical output 
is $1$, requiring the particle to occupy the right well. In a symmetric 
potential with no energy bias, however, there is no deterministic 
preference for either well. Consequently, at low noise the particle 
remains near its initial position; at high noise it switches between 
wells randomly with equal probability; and only at intermediate noise 
levels do a sufficient number of trajectories reach the desired right well 
to produce a measurable logical response. The system therefore possesses 
no intrinsic directional bias, and the logical operation relies almost 
entirely on fluctuations. An analogous argument applies to the AND gate 
(Fig.~\ref{fig:logical_response}~{\textcolor{blue}{(b)}}), where the same 
input combinations produce $I_{\mathrm{tot}} = 0$ but now require the 
output to be $0$, leading to a similar plateau structure near $c = 1$.}

\mdas{Now, we present the outcomes of our methodical analyses for the entropically asymmetric 
systems intended to perform as logic gates. In our current study,} it is observed that systems 
with greater width asymmetry 
\mdas{(decreasing order of $c$ for OR logic systems and increasing magnitude of $c$ for AND logic systems)} 
consistently outperform the symmetric case, 
yielding a higher \mdas{maximum value for} $P_{\mathrm{logic}}$ 
across the noise range considered. 
\mdas{This finding implies that it is possible to construct logic gates with a certain level of reliability if a high entropic bias between the two logical output states is implemented.}
However, even for the most asymmetric configuration, the 
system does not achieve a perfect logical response; $P_{\mathrm{logic}}$ 
remains below unity for all values of $D$.

We suggest that \mdas{although the perfect logical responses are not obtained from this class of 
systems with width asymmetry, the outcomes have fundamental importance. 
This is because here the left and the right potential wells of the 
double-well potential, i.e., the logical output states $0$ and $1$, maintain the same energy level 
to begin with, i.e., they are energetically equivalent for the unperturbed systems. 
\pcmd{In the case of} the related previous studies~\cite{Murali_prl_2009,Chen_pre_2026,Cai_2025} 
which report better logical responses, \pcmd{involve the description of the standard framework where the resulting underlying double-well potential has different well depths.} This indicates that for those cases,   
the two logical output states $0$ and $1$, designated by the left and the right wells of the 
potential, possess different energy values even when no input is applied to the system. This \pcmd{fact can be explained in a physical system with refering to the depiction of an analog circuit subject to disparate thresholds, or a DC bias, or both together considered in earlier studies~\cite{BULSARA2010424,Murali_apl_2009} \mdas{[Refs.]}. Their presence creates an effective asymmetric bistable system or a double-well potential with two distinct values of the energy minima. 
Therefore, a fundamental energetic asymmetry appears in the intrinsic potential due 
to the different depths of its two wells, leaving the intrinsic energy values of the binary memory states at different levels.} 
Our present analysis shows an alternative and potential path to forming  
logic systems with a moderate extent of reliability, even in situations 
when the output states are not inherently biased.} 

\mdas{We understand that the} width 
asymmetry alone is insufficient \mdas{to produce logic systems which can approach perfect 
($\sim$ 100\%) reliability}. Therefore, an additional energy bias \mdas{will be beneficial} to 
further improve the logical performance. \mdas{This fact} is investigated \mdas{in detail} 
\and{discussed} in the following subsection. 

\subsubsection*{Entropic and Energy Bias}

To investigate the combined effect of width asymmetry, \mdas{i.e., entropic bias} 
and energy bias on logical performance \mdas{of the systems}, 
we introduce both simultaneously. \mdas{This modifies the form of the potential $V(x)$ as}


\begin{figure*}[ht]
  \centering
    \centering
    \includegraphics[width=0.92\linewidth,height=4cm]{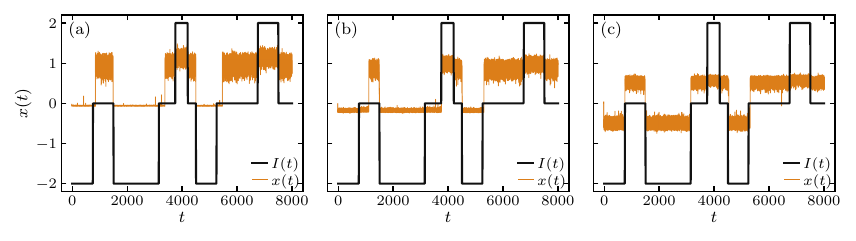}
  \vspace{0.1cm}
    \centering
    \includegraphics[width=0.92\linewidth,height=4cm]{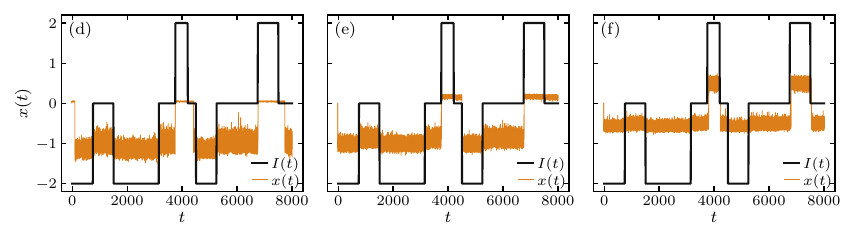}

    \caption{Input--output time series of the OR gate (top row, (a)--(c)) 
    and AND gate (bottom row, (d)--(f)) at fixed noise intensity $D = 0.125$, 
    shown for selected values of the width asymmetry parameter $c$ at their 
    respective optimal bias $b_{\mathrm{optimal}}$. Top row: $c = 0.08$ (a), 
    $c = 0.34$ (b), and $c = 1.00$ (c) for the OR gate. Bottom row: 
    $c = 1.92$ (d), $c = 1.66$ (e), and $c = 1.00$ (f) for the AND gate.}
  \label{fig:trajectories_bias}
\end{figure*}

\begin{equation}
    V(x) = h \left[ 1 + \left( \frac{2x}{S(x)} \right)^4 
    - 2\left( \frac{2x}{S(x)} \right)^2 \right] + bx
    \label{eq:potential_modi}
\end{equation}

\begin{figure}[ht]
\centering
\begin{minipage}{0.49\linewidth}
    \centering
    \includegraphics[width=\linewidth,height=3.5cm]{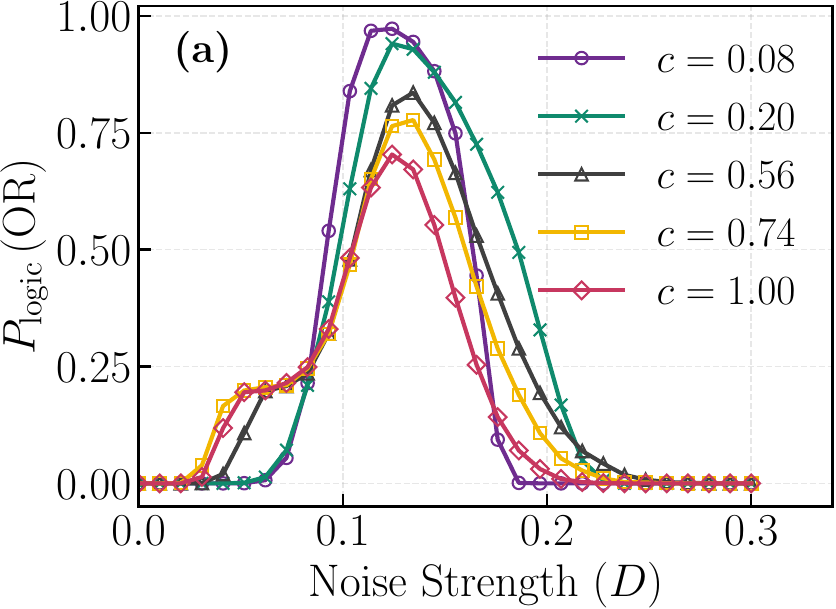}
\end{minipage}
\hfill
\begin{minipage}{0.49\linewidth}
    \centering
    \includegraphics[width=\linewidth,height=3.5cm]{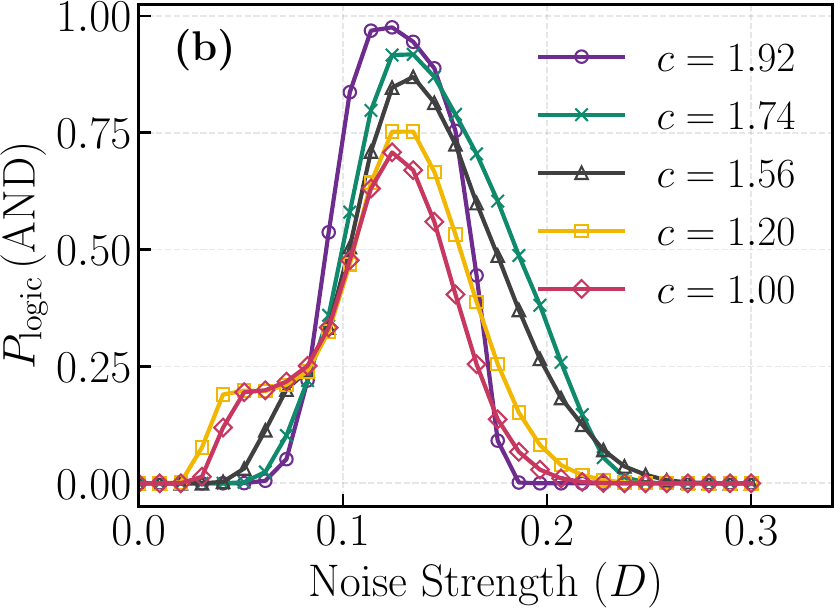}
\end{minipage}
\caption{Probability of logical success $P_{\mathrm{logic}}$ as a function 
of noise intensity $D$ for the (a) OR gate and (b) AND gate, evaluated 
over $N_{\mathrm{trial}} = 2 \times10^4$ trials. Results are shown for an 
asymmetric double-well potential with varying width asymmetry parameter 
$c$, in the presence of a fixed energy bias $b = 0.4$ (OR gate) and 
$b = -0.4$ (AND gate). The thermalization and measurement intervals are 
$T_{\mathrm{settle}} = T_{\mathrm{measure}} = 300$, with occupation 
threshold $\theta = 0.98$.}
\label{fig:logical_response_bias}
\end{figure}

Here, $b$ is an energy bias parameter that introduces a tilt in the potential, breaking the energetic equivalence of the two wells. A positive value ($b > 0$) favors the right well and is used for the OR gate, while a negative value ($b < 0$) favors the left well and 
is used for the AND gate. \textcolor{black}{The corresponding 
input--output time series for the case of combined entropic and 
energy bias are shown in Fig.~\ref{fig:trajectories_bias}}
\pcmd{for some representative scenarios when the systems exhibit optimized logical response. The results include both OR and AND logic systems.} 
\mdas{From the numerical simulation study with the above form of the potential,} 
it is found that neither bias \mdas{at their individual level} 
alone is sufficient to \mdas{produce a maximum of} $P_{\mathrm{logic}} \sim 1$. 
A combination of both is required to achieve \mdas{nearly perfect} logical response 
\mdas{in the systems.} Systems with greater width asymmetry attain a higher  
\mdas{optimized} $P_{\mathrm{logic}}$ at \mdas{the same magnitude of the energy bias} relative to 
the symmetric case as shown in Figs.~\ref{fig:logical_response_bias}~{\textcolor{blue}{(a)--(b)}}, 
\mdas{for the OR and AND logic systems, respectively}. 
\mdas{Moreover, further investigations reveal that a relatively weaker energy bias is sufficient 
to get maximized logical response as the degree of asymmetry increases in the system.} 
\mdas{The observations are the same for both types of systems performing as 
OR and AND logic gates.}

\begin{figure*}[ht]
  \centering
    \includegraphics[width=0.99\linewidth,height=4.7cm]{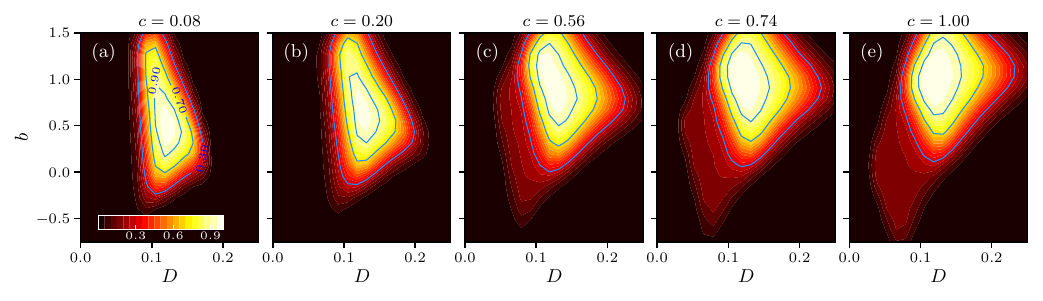}
    \label{fig:contour_OR}
    \includegraphics[width=0.99\linewidth,height=4.7cm]{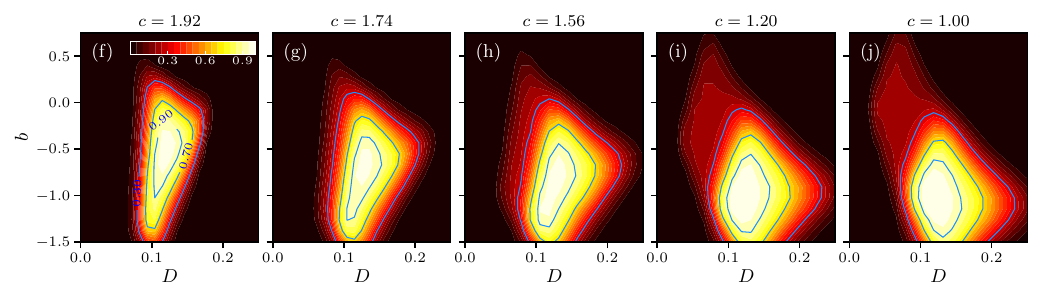}
    \label{fig:contour_AND}
    \caption{Contour plots of the logical success probability,
    $P_{\mathrm{logic}}$, in the $(D,b)$ plane.
    Panels (a)--(e) correspond to the OR gate, while panels (f)--(j)
    correspond to the AND gate, for different values of the width-asymmetry
    parameter $c$. The probabilities are estimated from
    $N_{\mathrm{trial}} = 10^4$ independent trials. The thermalization and
    measurement times are fixed at
    $T_{\mathrm{settle}} = T_{\mathrm{measure}} = 300$, and the occupation
    threshold is $\theta = 0.98$.}
  \label{fig:contour}
\end{figure*}

To provide a global picture of the logical performance, we present the contour 
plots of $P_{\mathrm{logic}}$ in the $(D,\, b)$ plane in 
Fig.~\ref{fig:contour}~{\textcolor{blue}{(a)--(e)}} for the OR gates and in Fig.~\ref{fig:contour}~{\textcolor{blue}{(f)--(j)}} for the  AND 
gates, computed for several values of the width asymmetry parameter $c$. 
These plots reveal the joint dependence of logical success on noise 
intensity and energy bias, and clearly show that increasing width 
asymmetry expands the region of high $P_{\mathrm{logic}}$ toward lower 
values of $b$.

\begin{figure}[ht]
\centering
\includegraphics[width=\linewidth,height=3.cm]{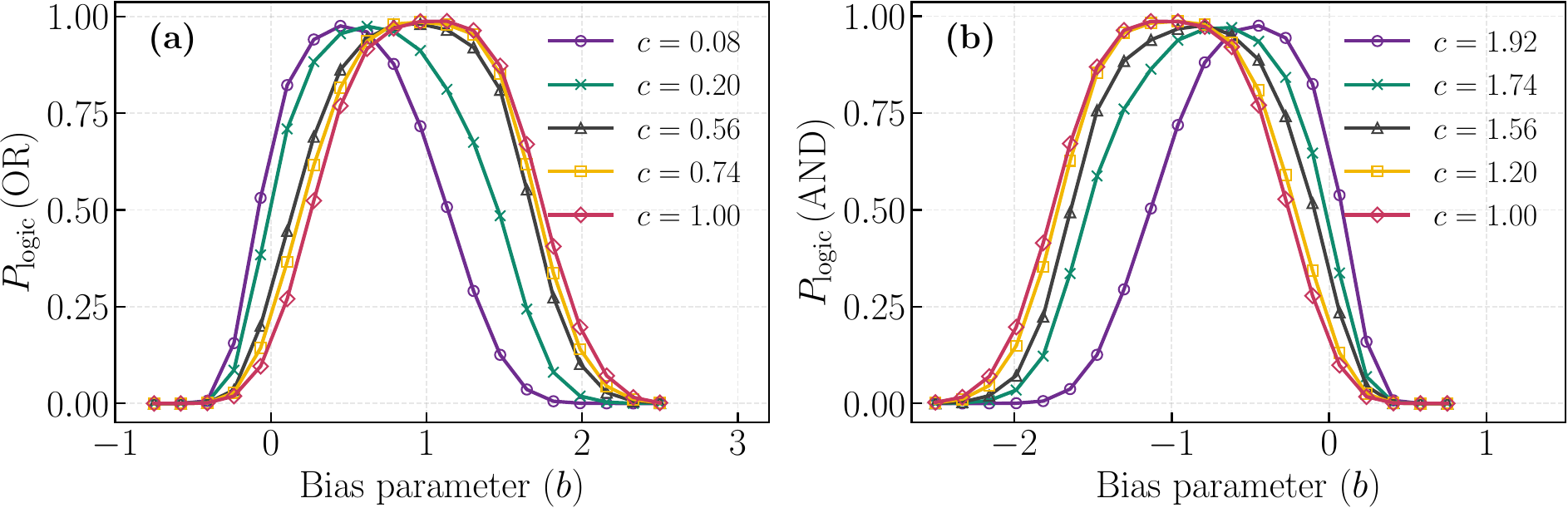}

\caption{Probability of logical success $P_{\mathrm{logic}}$ as a function
of energy bias $b$ for the (a) OR gate and (b) AND gate, at fixed noise
intensity $D = 0.125$, for several values of width asymmetry parameter
$c$. Results are evaluated over $N_{\mathrm{trial}} = 2 \times10^4$ trials, with
$T_{\mathrm{settle}} = T_{\mathrm{measure}} = 300$ and occupation threshold
$\theta = 0.98$.}

\label{fig:external_bias}
\end{figure}

To isolate the effect of energy bias, we fix the noise intensity at 
$D = 0.125$ and vary the external bias $b$. 
As shown in Fig.~\ref{fig:external_bias}\textcolor{blue}{(a)--(b)}, systems with greater width asymmetry 
attain a higher, \mdas{and also the maximized} $P_{\mathrm{logic}}$ at a smaller \mdas{modulus} 
value of $b$ \mdas{for both types of logic gates, OR and AND}.   
\mdas{This confirms} that entropic asymmetry reduces the \mdas{magnitude of the} energy bias 
required to achieve reliable logical operation. \mdas{Here, we mention one important point 
regarding the sign of the bias $b$, which is observed to develop the maximized logical response 
for two different types of logic gates, OR and AND. The results show that $b$ requires a 
positive sign for the construction of the reliable OR logic gates, according to the convention in 
our present study. $b$ is found to have a negative sign for the maximum value of 
$P_{\mathrm{logic}}$ for the systems exhibiting AND logical response. This is understandable as
the opposite types of bias favor the formation of the OR and AND logic systems.}

\textcolor{black}{It is further observed that 
the plateau in $P_{\mathrm{logic}}$ present in the symmetric case 
(Fig.~\ref{fig:logical_response}) disappears once a sufficient energy 
bias is applied. This can be understood as follows: the energy bias 
breaks the well symmetry and introduces a deterministic preference 
toward the correct output well, thereby resolving the \mdas{matter concerning the} 
ambiguous trajectories that were responsible for the plateau. 
Consequently, $P_{\mathrm{logic}}$ rises sharply with $b$ for the symmetric 
case, whereas asymmetric systems require a smaller bias to achieve 
the same effect. This is because the well-width asymmetry already provides a 
partial directional preference.}

\begin{figure}[ht]
\centering
\begin{minipage}{0.49\linewidth}
    \centering
    \includegraphics[width=\linewidth,height=3.5cm]{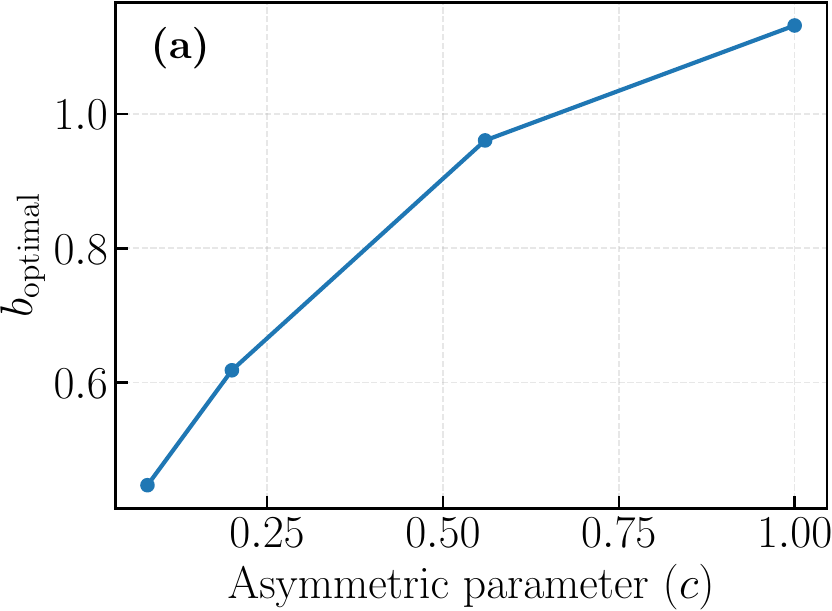}
\end{minipage}
\hfill
\begin{minipage}{0.49\linewidth}
    \centering
    \includegraphics[width=\linewidth,height=3.5cm]{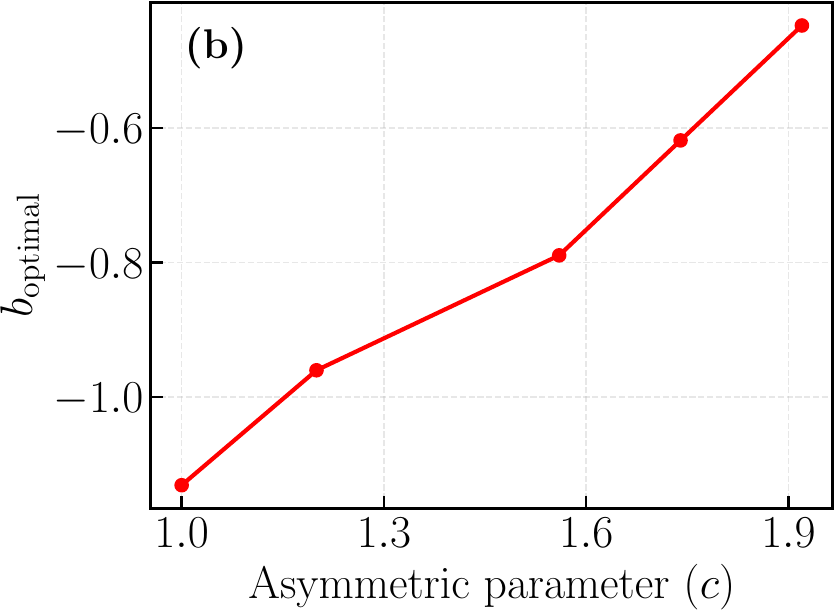}
\end{minipage}
\caption{Optimal energy bias $b_{\mathrm{optimal}}$ as a function of 
the width asymmetry parameter $c$ for the (a) OR gate and (b) AND gate, 
at fixed noise intensity $D = 0.125$. $b_{\mathrm{optimal}}$ is defined 
as the value of $b$ at which $P_{\mathrm{logic}}$ attains its maximum 
$P_{\mathrm{max}}$. Results are evaluated over $N_{\mathrm{trial}} = 20000$ 
trials, with $T_{\mathrm{settle}} = T_{\mathrm{measure}} = 300$ and 
occupation threshold $\theta = 0.98$.}
\label{fig:optimal}
\end{figure}

To further quantify this advantage, we extract the optimal bias 
$b_{\mathrm{optimal}}$, defined as the value of $b$ at which 
$P_{\mathrm{logic}}$ attains its maximum $P_{\mathrm{max}}$, for each 
value of the asymmetry parameter $c$. 
\mdas{This has been done for both types of logic gates, OR and AND. 
The results have been illustrated in Fig.~\ref{fig:optimal}~{\textcolor{blue}{(a)--(b)}}
for the OR and AND logic systems, respectively.}   
As it is inferred from the Fig.~\ref{fig:optimal}, 
\mdas{the modulus value of} $b_{\mathrm{optimal}}$ increases monotonically 
as $c$ approaches the symmetric case ($c = 1$) \mdas{for both kinds of logic gates.}  
This investigation validates that greater width asymmetry reduces 
the \mdas{requirement for the} energy bias needed to maximize logical performance. 

\section{Conclusion}







In this work, we have systematically investigated the role of width 
(entropic) asymmetry in a bistable double-well potential on the formation 
of OR and AND logic gates via logical stochastic resonance. 
\mdas{Previous significant studies have revealed that the introduction of depth or energetic asymmetry in the double-well potential model of the bistable system studied from the perspective of logical stochastic resonance improves the logical response of the system. However, the role of the width asymmetry in the potential, the other fundamental type of asymmetry that can be present in the systems, has not been thoroughly explored yet in this context.}

\mdas{Recently, the} width asymmetry has been shown to play important roles in 
noise-driven phenomena \mdas{related to barrier crossing}~\cite{innerbichler_2020}, including \mdas{significant dynamical processes like} memory erasure~\cite{Rai_vipul, 
Bechhoefer_prl} and dynamical hysteresis~\cite{ghosh2025dynamic}. 
The present study extends this \mdas{crucial} line of investigation to the domain of 
noise-assisted logic operations. \mdas{This is one of the similar important areas where a thorough understanding regarding the interplay between the nonlinearity in the systems, signal and environmental noise is essential. This interpretation can suggest advancement of the efficacy of the outcome of the process. With this objective,} we have performed a methodical investigation of how the degree of entropic asymmetry, controlled by the well-width parameter $c$, \mdas{influences the logical responses exhibited by the systems.}

A key feature of the width-asymmetric potential is that the two logical 
output states remain energetically equivalent in the unperturbed system, 
making the asymmetry purely geometric (entropic) in nature. 
\mdas{This feature is distinct from previously studied cases with energetic asymmetric systems, where the intrinsic logical output states are energetically biased to perform a given logical operation. For example, the left well and the right well minima of the underlying double-well potential, representing the logical states $0$ and $1$, are maintained at different energy levels. Which minimum will have higher or lower energy as compared to the other is decided by the fact that which logical outcome is expected from the system. It has been found in earlier studies that the opposite types of energetic asymmetry support the formation of the OR and AND logic gates. Here, we show that this inherent energetic non-equivalence in the intrinsic logical output states is not essential to construct logic gates of the given kinds. One can have the advantage of starting with the energetically equivalent output states to proceed with designing the planned logic gates.}

We find that entropic asymmetry alone is sufficient to build 
logic gates \mdas{of OR and AND type within the present 
framework considered here,} with a moderately reliable 
logical response \mdas{at an optimal range of noise strength}. \mdas{The opposite types of width asymmetry in the model double-well potential or entropic asymmetry in the system are required to be implemented to produce OR and AND logical responses. We also point out that with the alternative assignments of the logical output states, the formation of NOR and NAND gates is also possible within the given setup.} \mdas{Methodical analyses suggest that} increasing the degree of 
width asymmetry consistently improves the \mdas{maximum value of} $P_{\mathrm{logic}}$, \mdas{quantifying the correct logical outcomes,} relative to 
the symmetric case. However, width asymmetry alone is not sufficient to 
achieve a highly reliable logical response \mdas{at the optimal noise range}; a certain extent of energy 
bias is additionally required. 

When an energy bias is introduced alongside entropic asymmetry, the 
logical performance \mdas{of the systems} improves substantially. \mdas{In this case, the systems exhibit nearly perfect logical responses within the optimal noise value window}. Crucially, a system with 
a greater width asymmetry requires a smaller energy bias to attain the same 
level of logical success. This demonstrates that entropic asymmetry acts as 
a geometric resource that reduces the energetic cost of reliable logic 
gate operation. These results establish width asymmetry as a meaningful and practical 
design parameter for noise-assisted logic devices. 

\mdas{Finally, we highlight some of the cases where entropic asymmetry emerges in real scenarios. This is to assert that our present theoretical analyses can guide a systematic path in developing logic systems for practical applicability.}   Geometric asymmetry 
of the described kind arises naturally in a broad class of confined physical 
environments --- including microfluidic 
channels~\cite{yang2006microfluidic, Kreuter2013, QIAO20212194, 
XU2020382} and optical trapping 
setups~\cite{ciampini2021experimental, innerbichler_2020, Albay:18} --- 
where spatial confinement produces entropic barriers that govern particle 
transport. Recent experimental work has further demonstrated that shaping 
the potential barrier profile in optical tweezers experiments can 
substantially enhance Brownian escape rates~\cite{Chupeau_2020}, 
confirming that width-asymmetric potentials of the kind considered here 
are experimentally accessible. \textcolor{black}{Beyond colloidal and microfluidic 
systems, potential shaping has also been realized in MEMS devices, 
where the bistable potential landscape can be controlled by tuning 
electrostatic parameters. In this case, it has been shown that reshaping the 
potential barrier provides a reliable strategy to improve signal quality 
under varying noise conditions~\cite{MEMS_SR_2026}. Taken together, 
these experimental realizations across diverse physical platforms 
confirm that the well-width asymmetry, as a geometric property of the 
potential, \mdas{can} constitute a practically relevant and tunable design 
parameter for noise-assisted logical operations \mdas{in physical devices. The basics of the methodical guide towards the control of the logical outcomes through the extent of asymmetry of this kind in actual systems can be inferred from our present analyses.}}

\section*{Acknowledgments}
V.R. acknowledges IIT Mandi for a fellowship. 
M.D. thanks SERB (Project No. SRG/2022/000296), 
Department of Science and Technology, Government of India, 
and IIT Mandi (Seed Grant No. IITM/SG/MUD/91) for financial support.
The High Performance Computing Cluster facility and Param Himalaya Supercomputing 
facility managed by IIT Mandi are also acknowledged. The authors thank Dr. Harsh Soni for
insightful discussions. \\\\
\textbf{Author contributions:} M.D. conceptualized and designed the research problem.
V.R. executed the research work. V.R. and M.D. analyzed the results. 
V.R. and M.D. prepared and finalized the manuscript draft.\\\\
The authors express no conflicts of interest.

\bibliographystyle{apsrev4-2-titles} 
\bibliography{reference}

\end{document}